\documentclass[aps,prl,twocolumn,showpacs,a4paper,unsortedaddress,amsmath,amssymb,byrevtex]{revtex4}
\usepackage[latin1]{inputenc}
\usepackage{graphicx}
\usepackage{dcolumn}
\usepackage{bm}
\usepackage{hyperref}

\begin{document}

\preprint{ffuov/02-01}

\title{Oscillating chiral currents in nanotubes: a  route to nanoscale  magnetic test tubes.}
\author{C. J. Lambert}
\author{S. W. D. Bailey}
\affiliation{Department of Physics, Lancaster University, Lancaster,
LA1 4YB, U. K.}
\author{J. Cserti}
\affiliation{Department of Physics of Complex Systems,
E{\"o}tv{\"o}s University, H-1117 Budapest, Hungary}

\date{\today}

\begin{abstract}
With a view to optimising the design of carbon-nanotube (CNT)
windmills and to maximising the internal magnetic field generated by
chiral currents, we present analytical results for the group
velocity components of an electron flux through chiral carbon
nanotubes. Chiral currents are shown to exhibit a rich behaviour and
can even change sign and oscillate as the energy of the electrons is
increased. We find that the transverse velocity and associated
angular momentum of electrons is a maximum for non-metallic CNTs
with a chiral angle of 18$^o$. Such CNTs are therefore the optimal
choice for CNT windmills and also generate the largest internal
magnetic field for a given longitudinal current. For a longitudinal
current of order $10^{-4}$ amps, this field can be of order
$10^{-1}$Teslas, which is sufficient to produce interesting
spintronic effects and a significant contribution to the self
inductance.

\end{abstract}


\maketitle

Chiral nanotubes and nanowires are of interest for a range of
 properties associated with external magnetic fields
\cite{chiral1}, their potential for creating nanoscale inductors
\cite{louie2} and their suggested role as building blocks in chiral
nanomotors \cite{Tu2005}. Examples studied to date include chiral
carbon and BC2N nanotubes \cite{louie1, miya, yev},  BN nanotubes
\cite{tom}, Fe-filled carbon nanotubes (CNTs) \cite{FE} and chiral
single-wall gold nanotubes \cite{gold}.

Most recently, interest in chiral currents has been rekindled by
their potential to drive CNT nanometre-scale motors
\cite{Cummings2000,Ruoff2000,Kang2005}. Such motors benefit from low
inter-wall friction \cite{Kolmogorov2000} and a high tensile
strength \cite{Lourie1998}, which allow one to engineer complex
structures \cite{Forro2000,Collins2001,Bourlon2004}, including
nanoscale bearings \cite{Cummings2000,Subramanian2007}, rotors
\cite{Bourlon2004,Fennimore2003}, oscillators
\cite{Legoas2002,Zheng2002,Williams2002,Zhao2003}, switches
\cite{Deshpande2006}  and telescopes \cite{grace}. The helical
arrangement of the atoms in chiral CNTs can be exploited
\cite{Saito2001,Lovovik2003}to produce a Brownian ratchet effect
\cite{Tu2005}, rotational and translational motion driven by thermal
gradients \cite{Schoen2006,Barriero2008} and motion \cite{Kral2002}
induced by circularly polarized light. Recently a new drive
mechanism for CNT windmills was proposed \cite{Bailey2008}, based
upon the torque generated by a flux of electrons passing through a
chiral CNT. It was shown that under appropriate conditions, the
dominant contribution to this torque is proportional to the flux of
angular momentum carried by electrons moving in the corresponding
infinite chiral CNT. This provides a useful guide for the design of
CNT windmills, since a calculation of the angular momentum and
associated chiral currents carried by electrons in infinite chiral
nanotubes does not require the solution of a scattering problem.

The fact that large chiral currents occur in CNTs is at first sight
surprising, since early first-principles studies
\cite{louie1,louie2} suggested that chiral CNTs do not carry a
significant chiral current. In this paper we study the energy and
voltage dependence of chiral currents in CNTs and show that the
small chiral current found in \cite{louie1} is a consequence of the
metallic nature of the CNT studied. In contrast for non-metallic
CNTs, we predict much larger chiral currents. We find that the
energy dependence of chiral currents is surprisingly rich. Indeed
the transverse current components can even change sign and oscillate
as the energy of the electrons is increased. We demonstrate that the
presence of large chiral currents produces significant magnetic
fields of order 0.1 Teslas  within the volume of a CNT, thereby
providing a novel magnetic test tube, which could be used to
manipulate the magnetic moments of encapsulated magnetic molecules
or particles. This internal field produces a significant
contribution to the self inductance of the CNT, which must be added
to the more usual contribution associated with the external magnetic
field \cite{inductor}.

To define the velocity components of electrons in chiral CNTs, we
follow the notation of \cite{Saito1998}, which introduces the
lattice vectors  of the corresponding  infinite 2D graphene sheet,
defined by $\rm
\mathbf{a_1}=\left(\frac{\sqrt3}{2},\frac{1}{2}\right) a$ and $\rm
\mathbf{a_2}=\left(\frac{\sqrt3}{2},-\frac{1}{2}\right)a$, where
$a=\sqrt 3a_{c-c}$ and $a_{c-c}=1.44 \AA$ is the carbon-carbon bond
length. An $\rm(n,m)$ CNT, where $\rm(0 \le m \le n)$ are integers,
is then defined by a transverse chiral vector $\rm\mathbf{Ch} =
n\mathbf{a_1} + m\mathbf{a_2}$, which wraps around the CNT
circumference  and a longitudinal translation vector
$\rm\mathbf{T}$. We are interested in resolving electron velocities
along axes $\rm x$ and $\rm y$, which are parallel to the unit
vectors $\rm \mathbf{\hat{Ch}}$ and $\rm \mathbf{\hat{T}}$
respectively. The velocity components are given by $\rm \hbar v_{x}
= \partial E(\mathbf{k})/\partial k_{x} $ and $\rm \hbar v_{y} =
\partial E(\mathbf{k})/\partial k_{y} $, where $\rm k_{x} =
\mathbf{k}.\mathbf{\hat{Ch}}$, $\rm k_{y} =
\mathbf{k}.\mathbf{\hat{T}}$ and $\rm E(\mathbf{k})$ is the energy
dispersion relation. In the simplest Slater-Koster scheme, $\rm
E(\mathbf{k})$ takes the form
\begin{equation}
\rm E(\mathbf{k})=\gamma\vert 1 + exp(-i\mathbf{k}.\mathbf{a_1})
+exp(-i\mathbf{k}.\mathbf{a_2})\vert,\label{a3}
\end{equation}
where $\rm \gamma$ is the hopping integral.

 Each mini band possesses a continuous longitudinal
wave vector $\rm k_{y}$ and is labelled by a quantised value of $\rm
k_{x}$, given by
\begin{equation}
\rm k^q_{x}=\frac{2\pi q}{\vert\mathbf{{Ch}}\vert}, \,\,\, (q=1,
\dots N_{\rm hex}),\label{a5}
\end{equation}
where $\rm N_{\rm hex}$ is the number of hexagons in a CNT unit
cell. For a given choice of $\rm E$ and $\rm q$, equation (\ref{a3})
can be solved to yield two values of $\rm k_{y}$. One of these
values, which we denote $\rm k^+_{y}(q,E)$, corresponds to a
positive longitudinal velocity $\rm v_{y}(q,k^+_{y}(q,E)$. The other
value, which we denote $\rm k^-_{y}(q,E)$, corresponds to a negative
longitudinal velocity $\rm v_{y}(q,k^-_{y}(q,E)$. In what follows,
we refer to these electrons as "right-moving" and "left-moving"
respectively. We are interested in the transverse velocities of
right-moving electrons, which we denote by $\rm
v_{x}(q,k^+_{y}(q,E))$. Our aim is to compute the total transverse
velocity $\rm v^{(n,m)}_{x}(E)$ of all right-moving electrons of
energy $\rm E$, which in units of the Fermi velocity $\rm v_F$ is

\begin{equation}
\rm v^{(n,m)}_{x}(E) = \sum_q v_{x}(q,k^+_{y}(q,E))/v_F,\label{a6}
\end{equation}
where the sum is over all mini bands with real longitudinal wave
vectors of energy $\rm E$.

At low-enough energies, the wave vectors of mini band $\rm q$
[namely $\rm (k_{x}^q,k^+_{y}(q,E))$ and $\rm
(k_{x}^q,k^-_{y}(q,E))$] can be chosen to be close to the K point
$\rm \mathbf{K}$. Since $\rm E(\mathbf{k})$ is an even function of
$\rm \mathbf{k}$, there will be another pair of wave vectors in the
vicinity of the second K point $\rm -\mathbf{K}$, given by $\rm
(k^{q'}_{x},k^-_{y}(q',E))=(-k^q_{x},-k^+_{y}(q,E))$ and $\rm
(k^{q'}_{x},k^+_{y}(q',E))=-(k^q_{x},k^-_{y}(q,E))$, which possess
negative and positive longitudinal group velocities respectively.

The contribution to the sum in equation (\ref{a6}) from these two
mini bands is
\begin{multline}
\rm v_{x}(q,E)=v_{x}(q,k^+_{y}(q,E))+v_{x}(q',k^+_{y}(q',E))\\
\rm =v_{x}(q,k^+_{y}(q,E))+v_{x}(-q,-k^-_{y}(q,E))\\
\rm =v_{x}(q,k^+_{y}(q,E))-v_{x}(q,k^-_{y}(q,E))\label{a7}
\end{multline}

The last line in this expression is useful, because it allows us
to focus on the contributions from a single K point only. It also
demonstrates that a non-zero transverse velocity arises from
trigonal warping, since for a perfect Dirac cone, the right hand
side of equation (\ref{a7}) would vanish. This suggests that at
low energies, an analytical expression for  $ \rm v_{x}(q,E)$, can
be obtained by writing the electron wave vector in the form ${\rm
\mathbf{k}=\mathbf{K}+} \mbox{\boldmath$\eta$}$, where $\rm
\mathbf{K}=(0,1)4\pi/(3a)$ is a vector pointing from the origin to
a K-point and Taylor expanding $\rm E(\mathbf{k})$ as a power
series in $\rm \eta_{x}$ and $\rm \eta_{y}$, where $\rm
\eta_{x}=\mbox{\boldmath $\eta$}.\mathbf{\hat{Ch}}$ and $\rm
\eta_{y}=\mbox{\boldmath $\eta$}.\mathbf{\hat{T}}$. This expansion
is of the form
\begin{equation}
\rm E^2=\sum_{i,j=0}^\infty c_{ij}\eta_{x}^i\eta_{y}^j\label{a8}
\end{equation}
where the coefficients $\rm c_{ij}$ satisfy $\rm
c_{00}=c_{10}=c_{01}=c_{11}=0$ and $\rm
c_{02}=c_{20}=3\gamma^2/(4a^2)$, $\rm c_{21}=-c_{03}/3=27\gamma^2
a^3 mn(m+n)/16(n^2+m^2+nm)^{3/2} $, $ \rm c_{12}=3\sqrt{3} \gamma^2
a^3 (m-n)(2n+m)(2m+n)/16(n^2+m^2+nm)^{3/2}. $ Differentiating this
with respect to $\rm \eta_{x}$ and writing $\rm
(k^{q}_{x},k^\pm_{y}(q,E))=\mathbf{K}+ (\eta_{x},\eta^\pm_{y})$,
yields to order $\rm [\eta^\pm_{y}]^2][\eta_{x}]$,
\begin{multline}
\rm 2Ev_{x}(q,E)= c_{12}[(\eta^+_{y})^2-(\eta^-_{y})^2]\\
+\eta_{x}\{2c_{21}[\eta^+_{y}-\eta^-_{y}]
+2c_{22}[(\eta^+_{y})^2-(\eta^-_{y})^2]\}\label{a9}
\end{multline}

To compute $\rm \eta^\pm_{y}$ for fixed $\rm \eta_{x}$ and $\rm E$,
we consider two cases: The first case arises when $\rm \eta_{x}\ne
0$, in which case one obtains
\begin{equation}
\rm v_{x}(q,E)/v_F=\frac{2\eta_{x}c_{21}}{E}\sqrt{E^2/c_{02}\,
-\eta_{x}^2}\label{a11}
\end{equation}
The second case corresponds to $\rm \eta_{x}=0$. In this case,
equation (\ref{a8}) yields to lowest order, $\rm
E^2=c_{02}\eta_{y}^2+c_{03}\eta_{y}^3$, which after solving by
iteration and combining with equation (\ref{a9}) yields
\begin{equation}
\rm v_{x}(q,E)/v_F=-c_{03}c_{12}E^2/c_{02}^{5/2}\label{a10}
\end{equation}
where $\rm v_F=c_{02}^{1/2}/\hbar=\sqrt{3}\gamma/2a\hbar$ is the
Fermi velocity.

 Since
$\rm k_{x}=2\pi q/\vert \mathbf{Ch}\vert$ and $\rm
K_{x}=\mathbf{K}.\mathbf{\hat{Ch}}=2\pi (n-m)/{3\vert
\mathbf{Ch}\vert}$, the value of $\rm \eta_{x}$ for the
lowest-energy mini band is $ \rm \eta_{1}=2\pi(-X/3)/\vert
\mathbf{Ch}\vert \label{1},$  where $\rm X=1$ for $\rm n=m+1$, $\rm
n=m+4$, etc, whereas $\rm X=-1$ for $\rm n=m-1$, $\rm n=m+2$, $\rm
n=m+5$, etc and $\rm X=0$ for $\rm n=m$, $\rm n=m\pm 3$, etc. Values
of $\rm \eta_{x}$ for higher-energy mini bands are obtained from
$\rm \eta_1$ by adding or subtracting integer multiples of $\rm
2\pi/\vert \mathbf{Ch}$. For non-metallic CNTs, where $\rm X=\pm1$,
the value of $\rm \eta_{x}$ for the second mini band is $ \rm
\eta_{2}=2\pi(2X/3)/\vert \mathbf{Ch}\vert, $ whereas for metallic
CNTs, $\rm \eta_{2}=\pm 2\pi/\vert \mathbf{Ch}\vert$.

For $\rm \eta_j\ne 0 $ equation (\ref{a11}) yields for the
dimensionless transverse velocity associated with channel $\rm
\eta_j$ of a (n,m) CNT,

\begin{equation}
\rm v_{x}^{{(n,m)}}(\eta_j,E)=v_{x}(q,E)/v_F=
v_j\epsilon_j^{3/2}(\epsilon-\epsilon_j)^{1/2}/\epsilon
 \label{3},
\end{equation}
where
\begin{equation}
\rm v_j=3\sqrt{6} mn(n+m)/(n^2+m^2+mn)^{3/2}[{\rm sign\,of\,}\eta_j]
 \label{4}.
\end{equation}
In this expression, $\rm \epsilon=E/\gamma$ and $\epsilon_j$ is the
energy minimum of the jth mini band (in units of $\rm \gamma$),
given by $ \rm \epsilon_j=(\sqrt{3}/2)a\vert \eta_j\vert $.

\begin{figure}[h]
\centering
\begin{tabular}{cc}

\includegraphics[width=0.5\linewidth,clip=]{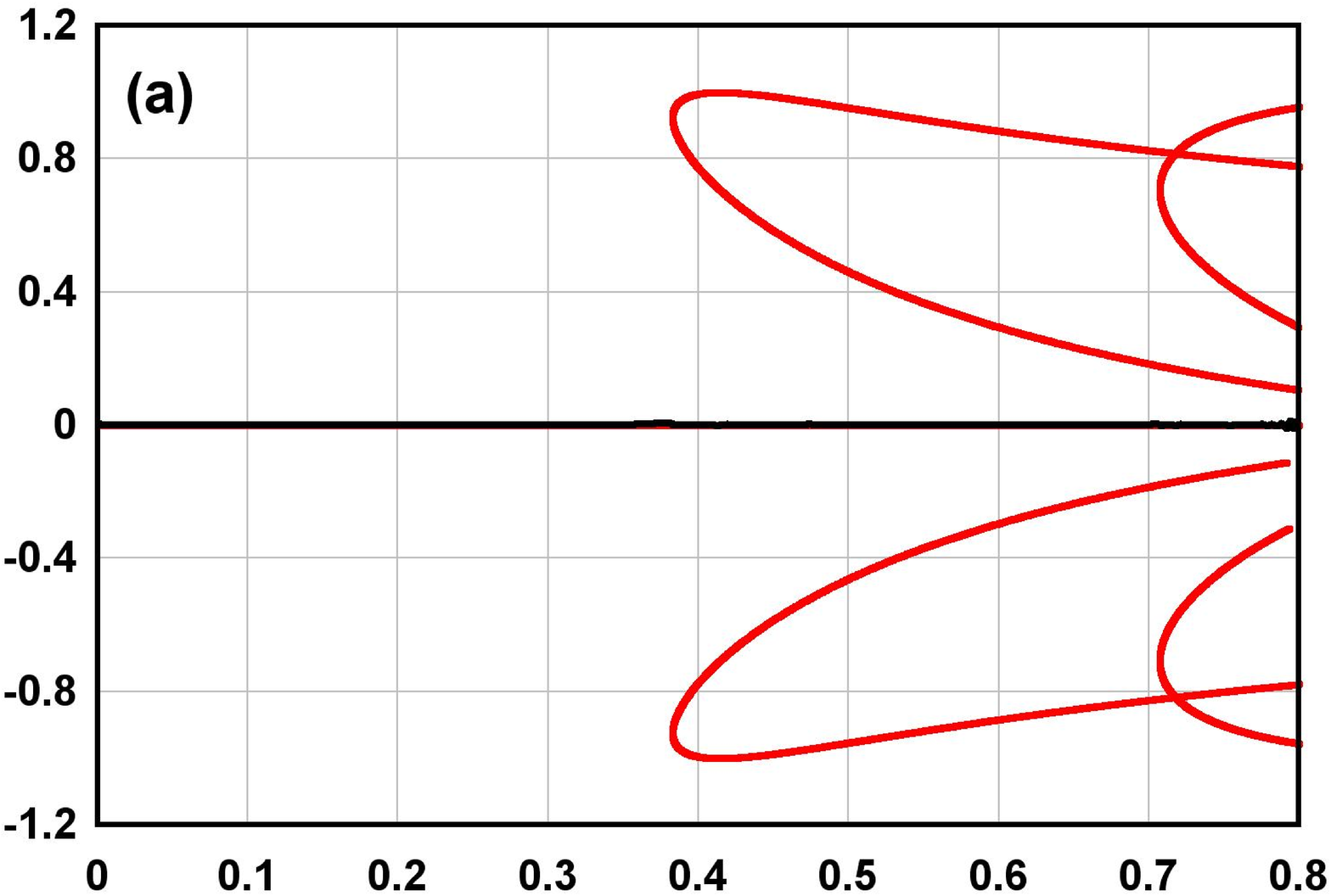}&

\includegraphics[width=0.5\linewidth,clip=]{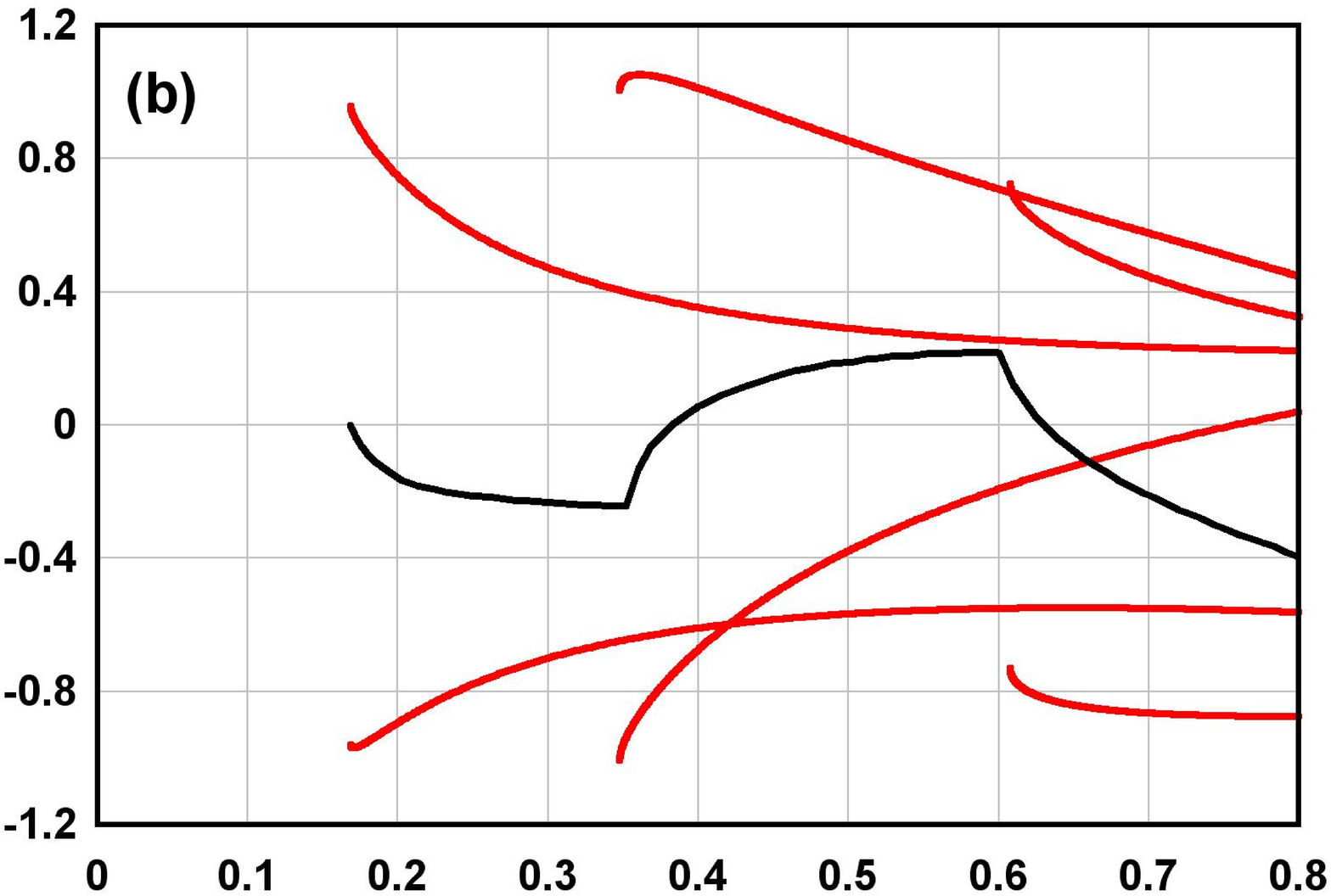} \\

\includegraphics[width=0.5\linewidth,clip=]{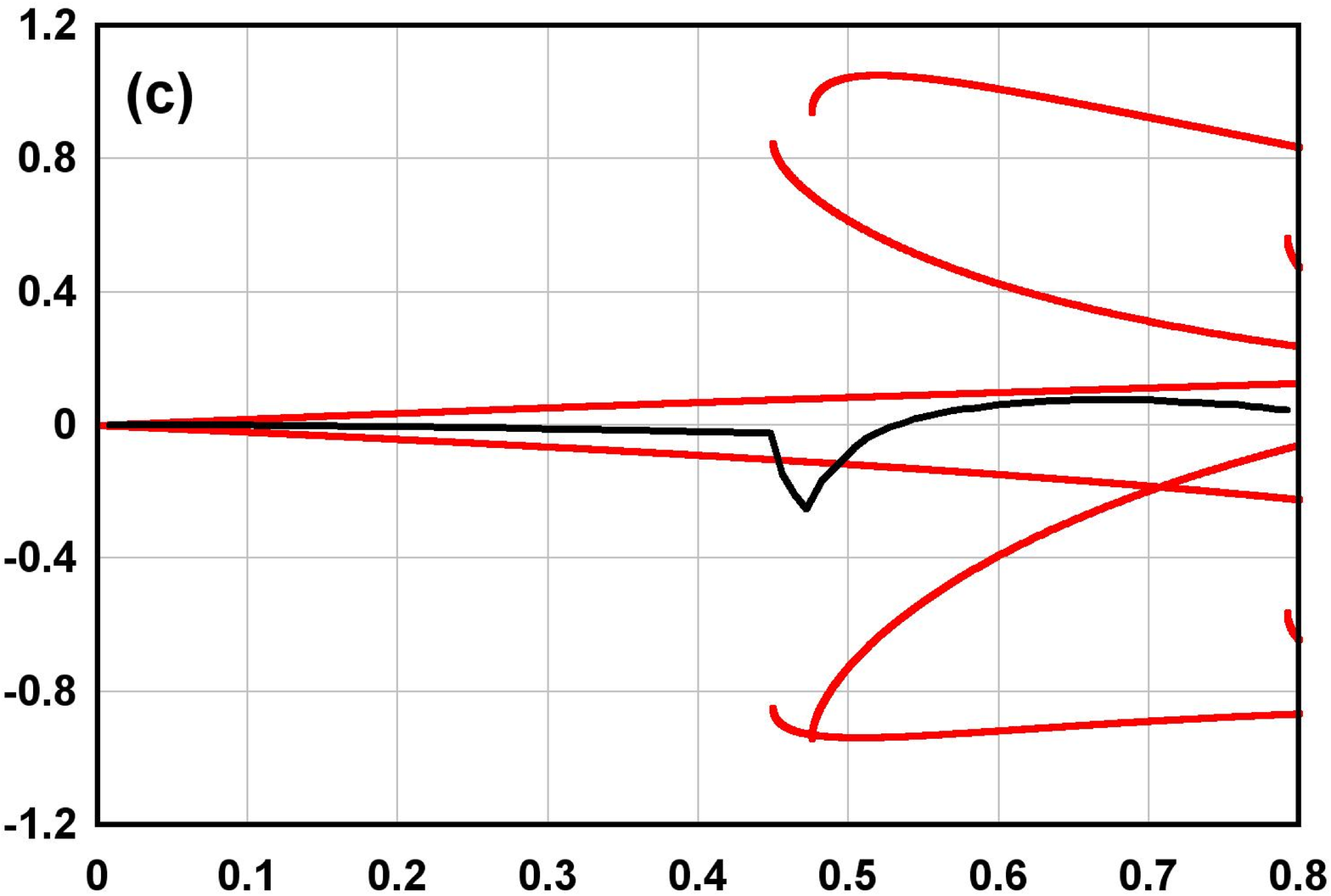} &

\includegraphics[width=0.5\linewidth,clip=]{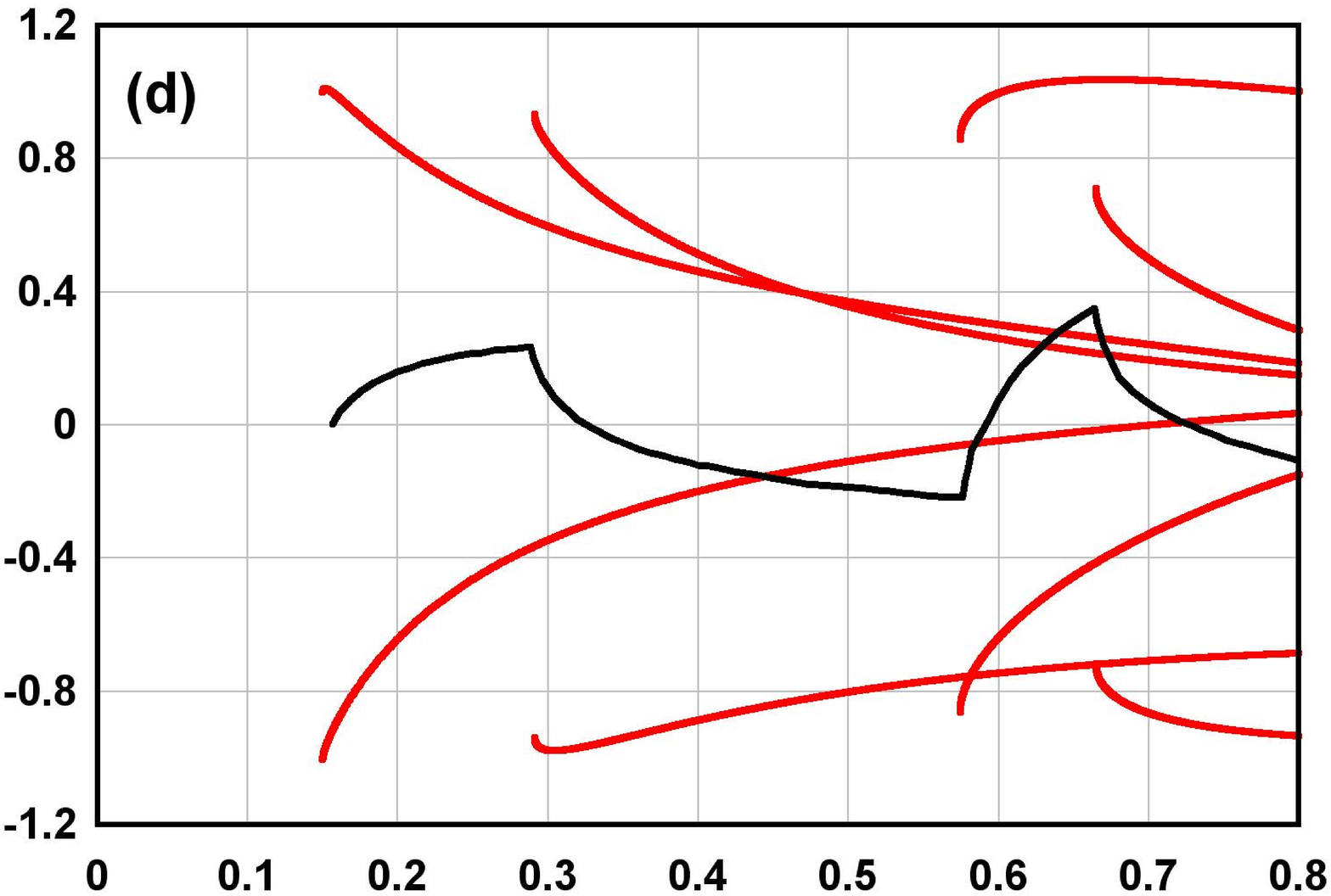} \\

\end{tabular}
\caption{Velocity components for the (8,m) family of CNTs plotted
against the energy $\rm E$ (eV), where m=8,4,5,6 in figs(a) to (d)
respectively. The red curves show the transverse velocities $\rm
v_{x}(q,k^+_{y}(q,E))/v_F$ of right-moving electrons belonging to
individual channels $\rm q$. The black curves show the total
velocity $\rm v^{(n,m)}_{x}(E)$ of Eq. (\ref {a6}). } \label{Fig4}
\end{figure}

The above result applies to all low-energy mini bands, except the
first mini band of metallic CNTs, for which $\rm \eta_1 =0$. In this
case, equation (\ref{a10}) yields

\begin{equation}
\rm v_{x}^{\rm{(n,m)}}(0,E)=v_0\epsilon^2
 \label{6},
\end{equation}
where
\begin{equation}
\rm v_0=\frac{\sqrt{3}mn((m^2-n^2)(2n+m)(2m+n)}{4(n^2+m^2+mn)^3}
 \label{7}.
\end{equation}
This quadratic dependence on $\epsilon$ means that low-energy
transverse currents in metallic CNTs is indeed small, in agreement
with \cite{louie1}. In contrast the square-root dependence arising
when $\rm \eta_j\ne 0$ means that transverse currents in
non-metallic CNTs are predicted to be much larger. This behaviour is
illustrated in the exact results of Fig.~\ref{Fig4}, obtained by
differentiating equation (\ref{a3}) with respect to the transverse
and longitudinal wave vectors. For each mini band $\rm q$, the red
curves of Fig.~\ref{Fig4} show the dimensionless velocities $\rm
v_{x}(q,k^+_{y}(q,E))/v_F$ as a function of $\rm E$ for the (8,m)
family of CNTs.  The black curves show the quantity $\rm
v_{x}^{{(n,m)}}(E)$, obtained by adding the values of the red curves
for each open channel of energy $\rm E$. As expected,
 for the achiral (8,8) CNT $\rm v_{x}^{(n,m)}(E)=0$, whereas for the chiral CNTs $ \rm
v_{x}^{(n,m)}(E)\neq 0$.

As well as predicting the energy dependence of transverse
velocities, equations (\ref{4}) and (\ref {7}) also yield the sign
of $\rm v_{x}^{\rm{(n,m)}}(\eta_j,E)$. For example, when $\rm n$ and
$\rm m$ are positive, the sign of $\rm v_1$ is equal to the sign of
(-$\rm X$) and therefore for the first open mini band of a (8,4)
CNT, $\rm v_{x}^{\rm{(8,4)}}(\eta_1,E)$ is negative, whereas for the
first mini band of a (8,6) CNT, $\rm v_{x}^{\rm{(8,6)}}(\eta_1,E)$
is positive. Similarly, when $\rm n\ge 0$ and $\rm m\ge 0$, the sign
of $\rm v_0$ is equal to the sign of $\rm (m-n)$. Hence equation
(\ref{7}) shows that for the lowest mini band of a (8,5) CNT, $\rm
v_{x}^{\rm{(10,4)}}(0,E) < 0$. For successive higher-energy mini
bands, $\eta_j$ alternates in sign and therefore equations (\ref{3})
and (\ref{4}) reveal that the transverse velocities of successive
higher-energy mini bands have a square-root $\rm \epsilon$
dependence, with an alternating sign. In Eq. (\ref{a6}), the label
$\rm q$ sums over $\rm N(E)$ right moving open channels, where $\rm
N(E)$ is a discontinuous function of $\rm E$, which changes by an
integer whenever new channels open or close. As predicted by
equation (\ref{a7}), the red curves in Fig.~\ref{Fig4}, show that
right-moving channels open or close in pairs and just as a pair of
channels open, their tangential velocities cancel. Consequently, as
shown in Fig.~\ref{Fig4}, $\rm v_{x}^{(n,m)}(E)$ is a continuous
function of $\rm E$, with a discontinuous first derivative.

For the purpose of designing a CNT windmill with the largest torque,
or a CNT with the largest internal magnetic field, it is of interest
to compute the maximum integrated flux of transverse momentum
carried by right-moving electrons in an infinite chiral CNT. Since
CNTs with $\rm \eta_1\ne 0$ possess the most favourable energy
dependence for $\rm v_{x}^{\rm{(n,m)}}(\eta_1,E)$, we focus on
non-metallic CNTs. Since $\rm v_{x}^{\rm{(n,m)}}(\eta_1,E)$ and $\rm
v_{x}^{\rm{(n,m)}}(\eta_2,E)$ have opposite signs, the sum $\rm
v_{x}^{\rm{(n,m)}}(\epsilon)=v_{x}^{\rm{(n,m)}}(\eta_1,E)+v_{x}^{\rm{(n,m)}}(\eta_2,E)$
increases monotonically with $\rm E$ for $\rm \epsilon_1\le
\epsilon\le\epsilon_2$ and then decreases with $\rm E$ for $\rm
\epsilon \ge\epsilon_2$, passing through zero when $\rm
\epsilon=\epsilon_{\rm max}$. Hence the maximum integrated flux of
transverse velocity is proportional to

\begin{equation}
\rm v^{{(n,m)}}_{\rm max}=\int_{\epsilon_1}^{\epsilon_{\rm
max}}d\epsilon v_{x}^{{(n,m)}}(\epsilon)
 \label{8},
\end{equation}
From equation (\ref{3}), one obtains $\epsilon_{\rm
max}=15\epsilon_{2}/14$ and

\begin{equation}
\rm v^{\rm{(n,m)}}_{\rm max}=\frac{\beta mn(n+m).[\rm{sign\, of}
\,(-X)]}{(n^2+m^2+mn)^{5/2}} \label{12}
\end{equation}
where
$
\beta=  2\sqrt{6}\,\pi^2\left[4
\arctan\left(\frac{1}{\sqrt{14}}\right)-
\arctan\left(\sqrt{\frac{8}{7}}\right)\right]\,
  \approx
10.9\,.
$

 Equation (\ref{12})
reveals that $\rm v^{{(n,m)}}_{\rm max}\rightarrow 0$ as $\rm
n,m\rightarrow \infty$, which reflects the fact that the angular
momentum carried by an electron wind is a consequence of the finite
diameter of the CNT and the finite difference between successive
values of $\rm \eta_j$. We also note that the optimum values of
(n,m), which maximise $\rm v^{{(n,m)}}_{\rm max}$ are those which
possess a chiral angle close to 
${\rm 60^o}\left(1-\frac{3}{\pi}\,\cos^{-1}\sqrt{11/20}\right)
\approx {\rm 18^o}$.

Having analysed the transverse velocity of electrons in a chiral
nanotubes, we now estimate the magnetic field generated by these
electrons. In what follows we assume that the chiral CNT can be
approximated by a long solenoid with a constant magnetic field $\rm
B$ inside the CNT and zero field outside. Since the number of
right-moving electrons per unit length  in channel $\rm q$ is $\rm
(1/\pi\hbar)dE/v_{y}(q,k^+_{y}(q,E))$, and these pass around the
circumference of the CNT in a time $\rm
\tau=|\mathbf{Ch}|/v_{x}(q,k^+_{y}(q,E))$, the contribution to the
tangential current per unit length from the $\rm q$th mini band is
$\rm [(1/\pi\hbar)dE/v_{y}(q,k^+_{y}(q,E))]e/\tau$. The magnetic
field $\rm B$ inside a solenoid is $\rm \mu_0$ multiplied by the
tangential current per unit length. Hence the field due to all
right-moving electrons in an energy window $\rm eV$ is
\begin{equation}
\rm B=\frac{2e\mu_0}{h|\mathbf{Ch}|}\int^{eV}_{0}dE
\sum_{q}[\frac{v_{x}(q,k^+_{y}(q,E))}{v_{y}(q,k^+_{y}(q,E))}]
\Theta(E-E_q)),
\end{equation}
where $\rm E_q$ is the lowest energy of the $\rm q$th mini band.
Since the current carried by these electrons is $ \rm
I=(1/\pi\hbar)\int^{eV}_{0}dE \sum_{q}\,\Theta(E-E_q))$, the
magnetic field can be written $ \rm B=\mu_0 I\alpha/|\mathbf{Ch}|$,
where
\begin{equation}
\rm \alpha=\frac{\int^{eV}_{0}dE
\sum_{q}[\frac{v_{x}(q,k^+_{x}(q,E))}{v_{y}(q,k^+_{y}(q,E))}]
\Theta(E-E_q))}{\int^{eV}_{0}dE \sum_{q}\Theta(E-E_q))}.
\end{equation}

The dimensionless parameter $\alpha$ is the average ratio of the
tranverse and longitudinal group velocities. At low voltages, Eq.
(\ref{a7}) allows this to be written as a sum over channels near the
K-point $\rm \mathbf{K}$ and using Eq.~(\ref{3}), one obtains
\begin{equation}
\rm \alpha=\frac{\int^{eV}_{0}d\epsilon
\sum_{j}[\frac{v_j\epsilon_j}{\sqrt{2}}]
\Theta(\epsilon-\epsilon_j)}{\int^{eV}_{0}d\epsilon
\sum_{j}\Theta(\epsilon-\epsilon_j)}.
\end{equation}
For $\epsilon_1 < \rm eV/\gamma<\epsilon_2$, this yields $\rm
\alpha=v_1\epsilon_1/\sqrt{2}$, which is of order unity for (8,4) or
(8,6) CNTs. Consequently, for a current of $10^{-4}$ amps, $\rm
\vert B\vert\approx 10^{-1}$ Teslas, which is large enough to
produce significant spintronic effects \cite{footnote}, such as
rotating the magnetic moment of a small magnetic island
\cite{Hortamani2007} or a metallocene encapsulated within a CNT
\cite{metallocene}. By computing the  energy stored in this magnetic
field, we find an associated inductance per unit length of $L =
\mu_0 \alpha^2/4\pi$. This internal field could be detected through
 NMR measurements on encapsulated spins and could form the
basis of scanning MR probe with nanometre spatial resolution.
Finally we note that if the longitudinal current is driven by an ac
voltage, the oscillations present in $\rm v^{(n,m)}_{x}(E)$ will
lead to the generation of higher harmonics, which may provide an
alternative probe of chiral currents.


\begin{thebibliography}{1111}
\bibitem{footnote}The corresponding magnetic flux $\phi = B |\mathbf{Ch}|^2/(4\pi)$
is given by $\phi/\phi_0=1.44\,{\sqrt 3}\,{\rm
x}\,(n^2+m^2+nm)^{1/2}\,I\alpha$. For a current of $10^{-4}$ amps,
$\phi/\phi_0\approx 10^{-4}\alpha$, which is too small to produce
significant orbital effects.
\end{thebibliography}
\end{document}